\documentclass[12pt]{article}
\usepackage{graphicx}
  \usepackage{amsthm,amsfonts}
  \usepackage{amsmath}
\newcommand{\bea}   {\begin{eqnarray}}
\newcommand{\eea}   {\end{eqnarray}}
\begin{document}
\renewcommand{\thefootnote}{\fnsymbol{footnote}}

\thispagestyle{empty}

\newcommand{\bphi} {\bar{\phi}}
\newcommand{\bpsi} {\bar{\psi}}
\newcommand{\bg} {\bar{g}}
\title{Four types of (super)conformal mechanics: \\D-module reps and invariant actions}

\author{N. L. Holanda\thanks{{\em e-mail: linneu@cbpf.br}}
\quad and\quad F.
Toppan\thanks{{\em e-mail: toppan@cbpf.br}}
\\
\\
}
\maketitle

\centerline{
{\it CBPF, Rua Dr. Xavier Sigaud 150, Urca,}}{\centerline {\it\quad
cep 22290-180, Rio de Janeiro (RJ), Brazil.}
~\\
\maketitle
\begin{abstract}
~\\

(Super)conformal mechanics in one dimension is induced by parabolic or hyperbolic/trigonometric
transformations, either homogeneous (for a scaling dimension $\lambda$) or inhomogeneous (at $\lambda=0$, with $\rho$ an inhomogeneity parameter).  Four types of
(super)conformal actions are thus obtained. With the exclusion of the homogeneous parabolic case, dimensional constants are present. \par
Both the inhomogeneity and the insertion of $\lambda$ generalize the construction of Papadopoulos [CQG 30 (2013) 075018; arXiv:1210.1719].\par
Inhomogeneous $D$-module reps are presented for the $d=1$ superconformal algebras $osp(1|2)$,
$sl(2|1)$, $B(1,1)$ and $A(1,1)$. For centerless superVirasoro algebras $D$-module reps are presented 
(in the homogeneous case for ${\cal N}=1,2,3,4$; in the inhomogeneous case for ${\cal N}=1,2,3$).\par
The four types of $d=1$ superconformal actions are derived for ${\cal N}=1,2,4$ systems.  When ${\cal N}=4$, the homogeneously-induced  actions are $D(2,1;\alpha)$-invariant ($\alpha$ is critically linked to $\lambda$); the inhomogeneously-induced actions are $A(1,1)$-invariant.\par 

 \end{abstract}
\vfill

\rightline{CBPF-NF-001/14}

\newpage
\section{Introduction}

In this paper we prove the existence of four types of conformally invariant actions for one-dimensional mechanical systems. In \cite{pap} Papadopoulos realized that under hyperbolic/trigonometric transformations of the fields, extra potential terms entered the conformal Lagrangians (these extra potential terms are not present when the standard, {\em parabolic}, realization of the conformal transformations is considered).\par
We generalize here the results of {\cite{pap}} in two distinct ways. 
At first we point out that a scaling dimension $\lambda$ is associated with the parabolic and hyperbolic/trigonometric $D$-module reps of the
conformal algebra $sl(2)$. In {\cite{pap}} $\lambda$ was only taken at the given fixed value which produces constant kinetic terms.  \par
The scaling parameter $\lambda$, however, cannot be so easily dismissed. In the supersymmetric generalizations (starting from the ${\cal N}=4$ supersymmetric extension) it acquires a critical property. Depending on the given supermultiplet, see e.g. formula (\ref{alphalambda}), it specifies under which of the exceptional $D(2,1;\alpha)$ (parametrized by $\alpha$) supersymmetry algebras, the system under consideration is superconformally invariant.\par  
Our second extension concerns the generalization to the {\em inhomogeneous} parabolic and hyperbolic/trigonometric conformal transformations of the fields (in \cite{pap} only homogeneous transformations were considered).
\par
We point out, see Appendix {\bf A}, the existence of two inequivalent classes of linear one-dimensional conformal transformations (and their supersymmetric extensions), the homogeneous ones, depending on the critical scaling $\lambda$, and the inhomogeneous ones, which are parametrized by the constant $\rho$.\par
Hyperbolic versus trigonometric transformations are mutually recovered via an analytic continuation. The passage from parabolic to hyperbolic transformations, see e.g. formula (\ref{parhypchange}), requires a singular change of variable. Under this change of variable the properties of their respective $D$-module reps (the scaling $\lambda$ or, in the inhomogeneous case, the parameter $\rho$) are easily recovered. The singularity of the change of variable is, on the other hand, responsible for the appearance in the Lagrangians of the extra potential terms that we mentioned before.  \par
On algebraic grounds the crucial difference between the hyperbolic and the parabolic $sl(2)$ transformations is the following. In the parabolic case the operator proportional to a time-derivative (the ``Hamiltonian") is given by the (positive or negative) $sl(2)$ root,
while in the hyperbolic case this Hamiltonian operator is associated with the $sl(2)$ Cartan generator.
This is the reason why, when we consider superalgebra extensions, the parabolic systems are supersymmetric in the ordinary sense, while the hyperbolic systems,
despite being superconformally invariant, are not ordinary supersymmetric theories (see the discussion in Appendix {\bf D}).\par
We end up, for one-dimensional conformal systems and their supersymmetric extensions, with four types of $D$-module reps and their associated (super)conformally invariant actions, namely the {\em homogeneous parabolic}, {\em inhomogeneous parabolic}, {\em homogeneous hyperbolic/trigonometric} and {\em inhomogeneous hyperbolic/trigonometric} cases.\par
Only for the very special homogeneous parabolic case the conformally invariant actions are based on power-law and contain no dimensional parameter. In all remaining cases we have at disposal at least one dimensional constant to play with.\par
In the following we present all four types of (super)conformal actions in various exemplifying $d=1$ situations: the $sl(2)$-invariant actions of a single boson, the $osp(1|2)$-invariant ($sl(2|1)$-invariant) actions of an ${\cal N}=1$ (${\cal N}=2$) supermultiplet. For
a given set of ${\cal N}=4$ supermultiplets, the actions are $D(2,1;\alpha)$-invariant in the two homogeneous cases and $A(1,1)$-invariant in the two inhomogeneous cases. The non-trivial relation
between homogeneous and inhomogeneous actions (discussed in Appendix {\bf C}) can be appreciated 
in a different basis obtained through non-linear field redefinitions. In the new basis the actions possess a constant kinetic term (plus interacting potentials), while the superconformal algebra is realized non-linearly.\par
The construction is also applied, in the Lagrangian setting, to (super)conformal actions in $d=2$ dimensions. The invariance in this case
is under a single copy (for the chiral models) or the direct sum of two copies (the full conformal invariance) of the Witt algebra (the centerless Virasoro algebra) and its supersymmetric extensions. \par 
Contrary to the $d=1$ case, in $d=2$, hyperbolic and parabolic active transformations of the field(s) produce the same conformally invariant output. As an example,
the inhomogeneous transformations applied on a single boson induce, see (\ref{d2liou}), the conformally invariant Liouville action, while the homogeneous
transformations induce a power-law conformally invariant action, see formula (\ref{d2hom}). These two-dimensional actions are related by a non-linear field redefinition.\par
From the point of view of representation theory we extend here in two directions the results of \cite{kuto} and \cite{khto} on $D$-module reps of finite $d=1$ superconformal algebras. We enlarge the $D$-module reps of the $osp(1|2)$, $sl(2|1)$, $B(1,1)=osp(3|2)$ and $A(1,1)=sl(2|2)/{\mathbb Z}$ superalgebras to the class of {\em inhomogeneous} ($\rho$-dependent) $D$-module reps, see (\ref{inhomrepsusy}).\par
We further construct the $D$-module reps of the ${\cal N}=1,2,3,4$ centerless superVirasoro algebras, both in the homogeneous case
(they are summarized in (\ref{homrepsvir})) and, for ${\cal N}=1,2,3$, inhomogeneous case (these results are summarized in (\ref{inhomrepsvir})). The explicit construction of these $D$-module reps is presented
in Appendix {\bf B}.\par
Conformal mechanics based on the $sl(2)$ algebra has been investigated since the work of de Alfaro, Fubini and Furlan \cite{dff}.
Models of superconformal mechanics have been presented in \cite{fr}--\cite{pap2} (for an updated review on superconformal mechanics and a list of recent references see, e.g., \cite{fil3}). For superconformal actions with oscillator potentials see also \cite{di} and \cite{bk}. For non-linear realizations see \cite{aku}. There are several reasons to study one-dimensional superconformal mechanics (more on that in the Conclusions). Here it is sufficient to mention the applications to test particles moving in the proximity of the horizon of certain black holes, see \cite{bmsv}.
In \cite{kuto} and \cite{khto} it was advocated the point of view that superconformal mechanics, in the Lagrangian setting, could be derived from the $D$-module reps of superconformal algebras. In most of the papers in the literature and all works cited in the \cite{fil3} review, the superconformal actions are based
on power laws, being dependent only on dimensionless constants (apart the optional addition of oscillatorial terms, what is known as the DFF trick \cite{dff}). This is what to be expected for the homogeneous parabolic $D$-module reps. The possibilities offered by the three remaining types of $D$-module reps (presenting dimensional constants), on the other hand, greatly enlarge the class of available superconformal systems. One should confront, for instance, the power law ${\cal N}=4$ superconformal systems with $A(1,1)$ or $D(2,1;\alpha)$ invariance investigated in \cite{{dprst},{bko}} and \cite{ikl0}--\cite{kl}, respectively, with the ${\cal N}=4$ actions presented in Section {\bf 9}. 
\par 
The scheme of the paper is as follows: in Section {\bf 2} we introduce the {\em homogeneous} parabolic and hyperbolic $D$-module reps of the Witt algebra and its $sl(2)$ subalgebra. The {\em inhomogeneous} $D$-module reps of the $sl(2)$ and Witt algebras are discussed in Section {\bf 3}.
In Section {\bf 4} we derive the different types of conformally invariant actions for a single boson. In Section {\bf 5} we extend the analysis to the bosonic, conformally invariant actions in $d=2$ dimensions. In Section {\bf 6} we collect the main properties of the finite $d=1$ superconformal algebras (together with their known $D$-module reps) and of the ${\cal N}=1,2,3,4$ centerless superVirasoro algebras. The new results on $D$-module reps for the finite $d=1$ superconformal algebras and for the centerless superVirasoro algebras are summarized in Section {\bf 7}. Different types of ${\cal N}=1,2$ superconformal actions in $d=1$ and supersymmetric chiral actions in $d=2$ are given in Section {\bf 8}. In Section {\bf 9} we present the four types of ${\cal N}=4$ superconformally invariant actions associated with the class of $(1,4,3)$ supermultiplets. In the Conclusions we point out the possible applications of our results and the future lines of investigations. The paper is complemented by four Appendices. A discussion about homogeneous versus inhomogeneous $D$-module reps is given in Appendix {\bf A}. In Appendix {\bf B} we present the explicit construction of the new supersymmetric $D$-module reps discussed in the main text. In Appendix {\bf C} 
the ${\cal N}=4$ superconformally invariant actions of Section {\bf 9} are presented in a new basis. In this basis the kinetic term is constant, while the superconformal algebra is realized non-linearly. In Appendix {\bf D} we point out that the one-dimensional, superconformally invariant, hyperbolic and trigonometric systems are not ordinary supersymmetric theories. 
  
\section{The bosonic case: homogeneous $D$-module reps of the $sl(2)$ and $Witt$ algebras}

The $sl(2)$ algebra is the conformal algebra in $d=1$ dimensions. Its three generators $D,H,K$ satisfy the commutation relations
\bea\label{sl2}
\relax [D,H]&=& H,\nonumber\\
\relax [D,K]&=&-K,\nonumber\\
\relax [H,K]&=& 2D.
\eea
The Cartan generator $D$ is the dilatation operator.\par
 A (parabolic) $D$-module representation of 
(\ref{sl2}) is given by the differential operators (depending on a single variable $t$ which, in application to physics, plays the role of time)
\bea
H&=&\partial_t,\nonumber\\
D&=&-t\partial_t-\lambda,\nonumber\\
K&=&-t^2\partial_t-2\lambda t.
\eea
The constant $\lambda$ is the scaling parameter. The above $D$-module rep is non-critical because the commutators (\ref{sl2}) close for any value of $\lambda$.\par

The Virasoro algebra {\em Vir} is the central extension of the algebra of one-dimensional diffeomorphisms (known as
``Witt algebra"). Its infinite generators $L_n$ ($n\in {\mathbb Z}$) satisfy the commutation relations
\bea\label{vir}
[L_m,L_n]&=& (m-n)L_{m+n}+ \frac{c}{12}m(m^2-1)\delta_{m+n,0}.
\eea

The Virasoro algebra contains $sl(2)$ as a subalgebra. It is obtained by restricting $n=\pm 1, 0$.\par
In the centerless  case the Witt algebra admits a parabolic $D$-module rep. Indeed
\bea
L^{par.}_n&=& -t^{n+1}\partial_t-\lambda_nt^n
\eea
give the commutators (\ref{vir}) with $c=0$ provided that the $\lambda_m$'s satisfy the set of equations
\bea
m\lambda_m-n\lambda_n&=& (m-n)\lambda_{m+n}.
\eea
A solution is recovered for
\bea
\lambda_n &=& n{\tilde\lambda}+{\tilde \gamma},
\eea
with ${\tilde\lambda}, {\tilde\gamma}$ arbitrary constants.\par
For the $sl(2)$ generators we obtain, in particular,
\bea
L^{par.}_{-1} &=& -\partial_t+({\tilde \lambda}-{\tilde\gamma})\frac{1}{t},\nonumber\\
L^{par.}_0&=& -t\partial_t-{\tilde \gamma},\nonumber\\
L^{par.}_1&=& -t^2\partial_t-({\tilde \lambda}+{\tilde \gamma})t.
\eea
The special value ${\tilde \gamma}={\tilde \lambda}$ allows us to identify, for $\lambda={\tilde\lambda}$,
\bea
L^{par.}_{-1}&\equiv&-H,\nonumber\\
L^{par.}_0&\equiv& D,\nonumber\\
L^{par.}_1&\equiv& K.
\eea
At this special value of ${\tilde\gamma}$ one of the root generators of $sl(2)$ is proportional to a time-derivative and,
in physics, can be identified with the Hamiltonian.\par
The constant ${\tilde\gamma}$ is arbitrary and can be changed via a similarity transformation. Indeed,
for $f(t)= sgn(t){\hat\gamma}ln|t|$, we have
\bea
L^{par.}_n&\mapsto& L^f_n= e^fL^{par.}_ne^{-f}=L^{par.}_n+{\hat\gamma}t^n.\nonumber
\eea
Therefore, ${\tilde\gamma}\mapsto {\tilde\gamma}-{\hat\gamma}$.\par
For the special choice ${\tilde\gamma}={\tilde\lambda}$, the parabolic $D$-module rep of the Witt algebra is
\bea
L^{par.}_n&=& -t^{n+1}\partial_t-(n+1){\tilde\lambda}t^n.
\eea
By using hyperbolic/trigonometric functions, a hyperbolic/trigonometric $D$-module rep of the Witt algebra
can be given. In the hyperbolic case the $c=0$ commutators (\ref{vir}) are satisfied for
\bea
L^{hyp.}_n&=& -\frac{1}{\mu}e^{n\mu\tau}(\partial_\tau+{\overline \lambda}_n),
\eea
if ${\overline\lambda}_n=n{\overline\lambda} +{\overline\gamma}$. The dimensional constant $\mu$ has 
been introduced here for dimensional reasons. Without loss of generality we can fix it at the $\mu=1$ value. In most of the cases, nevertheless, it is convenient to explicitly keep it in the equations in order to facilitate a dimensional analysis.\par
The $sl(2)$ generators read now
\bea
L^{hyp.}_1 &=& -\frac{1}{\mu}e^{\mu\tau}(\partial_\tau +{\overline\lambda}+{\overline\gamma}),\nonumber\\
L^{hyp.}_0&=& -\frac{1}{\mu}(\partial_\tau+{\overline\gamma}),\nonumber\\
L^{hyp.}_{-1}&=& -\frac{1}{\mu}e^{-\mu\tau}(\partial_\tau-{\overline\lambda}+{\overline\gamma}).
\eea
In the hyperbolic case the generator proportional to the time-derivative (the ``Hamiltonian") coincides,
for ${\overline\gamma}=0$, with the $sl(2)$ Cartan generator $L^{hyp.}_0$.\\ Just like the parabolic case,
the constant $\overline\gamma$ can be shifted by a similarity transformation.\par
At this point it is important to stress that the parabolic and the hyperbolic $D$-module reps of the Witt algebras are singled
out, among the most general class of $D$-module reps, by the aforementioned very special property. 
Namely, that for a specific value
of the constant parameter (either ${\tilde\gamma} $ or ${\overline\gamma}$), one of the $sl(2)$ generators is proportional to the Hamiltonian. The mathematical difference between the parabolic and the hyperbolic $D$-module reps can be stated as follows. In the parabolic case, the Hamiltonian is identified with the 
(positive or negative) $sl(2)$
root generator while, in the hyperbolic case, the Hamiltonian is identified with the $sl(2)$
Cartan generator. This difference proves to be crucial in the construction of conformally invariant actions.\par
From an algebraic point of view the hyperbolic $D$-module rep can be recovered from the parabolic $D$-module rep via a singular transformation. Let us call, for simplicity, ${\overline L}_n= L^{hyp.}_n$ when we fix
the
values $\mu=1$ and $\overline\gamma=0$. Therefore
${\overline L}_n= -e^{n\tau}(\partial_\tau+n{\overline\lambda})$.
For $t>0$ the change of variable 
\bea\label{parhypchange}
&t\mapsto \tau(t)=ln(t)&
\eea
allows to recover the
parabolic rep ${\overline L}_n=-t^{n+1}\partial_t-n{\overline\lambda}t^n$ at the specific values, for its constants, ${\tilde\lambda}=
{\overline\lambda}$ and ${\tilde\gamma}=0$.\par
The 
\bea\label{scalingequality}
{\tilde\lambda}&=&{\overline\lambda}
\eea
relation is of particular importance. Extended to superconformal algebras with ${\cal N}\geq 4$ (the ones,
as discussed in the Introduction, where the criticality of the scale parameter plays a role), it implies that
the same critical scaling is recovered in both parabolic and hyperbolic cases
(we will see this property at work in the following of the paper).\par
The singularity of the transformation connecting parabolic and hyperbolic $D$-module reps has the consequence, for the respective conformal invariant actions, that they are not (at least trivially) related. 
With respect to the parabolic case, in the hyperbolic case extra potential terms appear due to the presence of the dimensional constant
$\mu$ (and due to the different identification of the Hamiltonian operator with the given $sl(2)$ generator).\par
The connection of the trigonometric case (that we do not need here to write down explicitly) with the hyperbolic case is simply given by an analytic continuation. One can perform a Wick rotation of the time coordinate
$\tau$ by identifying a new periodic variable $\theta$ ($\tau\equiv i\theta$). Alternatively, the analytic continuation can also be obtained by performing a Wick rotation of the dimensional constant $\mu$, mapping $\mu\mapsto i \mu$. It will be shown in the following that the extra potential terms entering the conformally invariant actions in the hyperbolic case are not bounded below, due to a ``wrong" sign.
Since they are proportional to $\mu^2$, the correct sign can be recovered through the latter Wick rotation.
The conformally invariant actions based on the trigonometric $D$-module transformations have therefore
well-defined, bounded from below, potentials.\par
As recalled in \cite{pap}, the group of diffeomorphisms $Diff({\mathbb R})$
of the real line induced by the hyperbolic $D$-module rep is promoted, in the trigonometric case,  to the group of diffeomorphisms $Diff({\bf S}^1)$ of the ${\bf S}^1$ circle.

\section{Inhomogeneous $D$-module reps of the $sl(2)$ and Witt algebras}

Besides distinguishing Witt algebra's $D$-module reps into the two classes of {\em parabolic} versus {\em hyperbolic/trigonometric} representations, another
discrimination can be introduced. It concerns the {\em homogeneous} versus the {\em inhomogeneous} representations.\par
Let $\varphi(t)$ be a time-dependent field. In the homogeneous case, the action of the Witt generators
is written down as
\bea\label{hom}
L_n (\varphi )&=& a_n{\dot\varphi} +b_n\varphi.
\eea
In the inhomogeneous case the generators act  as
\bea\label{inhom}
L_n (\varphi )&=& a_n{\dot \varphi} + d_n.
\eea
In both cases the closure of the $c=0$ (\ref{vir}) commutators is guaranteed, provided that 
the coefficients $a_n, b_n$ and $a_n, d_n$ are fixed to proper values (the coefficients $b_n, d_n$ coincide; for clarity reasons in application to conformal actions, it will be however convenient to denote with different letters their respective normalization constants).\par
The parabolic subcase requires
\bea\label{par}
a_n=-t^{n+1}, && b_n={\tilde\lambda}{\dot{a}_n}, \quad\quad d_n={\tilde\rho}{\dot{a}_n}.
\eea

The hyperbolic subcase requires
\bea\label{hyp}
a_n= -\frac{1}{\mu}e^{n\mu \tau}, && b_n= {\overline{\lambda}}{\dot{a}_n}, \quad\quad d_n={\overline\rho}{\dot{a}_n}.
\eea
 
Taking into account the discussion in Appendix {\bf A},
the overall result is the existence of four types of $D$-module representations of the Witt algebra,
labelled as follows:\par
~
\par
~~~{\em I} ({\em Hom. par.}) - the {\em homogeneous parabolic} rep,\par
~~{\em II} ({\em Inh. par.}) - the {\em inhomogeneous parabolic} rep,\par
~{\em III} ({\em Hom. hyp.}) - the {\em homogeneous hyperbolic} rep,\par
~{\em IV} ({\em Inh. hyp.}) - the {\em inhomogeneous hyperbolic} rep.\par

~\par
Let $[t]=[\tau]=-1$ be the scaling dimension of the time coordinate(s) 
 (therefore $ [\mu]=1$). Let us furthemore set the scaling dimension of the field
$\varphi$ being given by $[\varphi] = s$.\par
For consistency, in the respective cases, the scaling dimensions of the ${\tilde\lambda}$,
${\tilde\rho}$, ${\overline\lambda}$, ${\overline\rho}$ parameters
are
\bea\label{scalings}
\relax &{ I}: [{\tilde \lambda}]=0, \quad\quad
{ II}:[{\tilde\rho}]= s,\quad\quad
{ III}: [\overline\lambda]=0,\quad\quad
{IV}: [\overline\rho]=s.&
\eea
For $s\neq 0$ the {\em Hom. par.} rep contains no dimensional parameter, while one dimensional parameter
(${\tilde\rho}$) is found in the {\em Inh. par.} rep, one dimensional parameter ($\mu$) in the
{\em Hom. hyp.} rep and two dimensional parameters ($\mu, {\overline\rho}$) in the {\em Inh. hyp.} rep.  \par
Similarly to the homogeneous case, the change of variable (\ref{parhypchange}) allows to connect the inhomogeneous parabolic and hyperbolic $D$-module reps. Under this transformation the relation
\bea
{\tilde\rho}&=&{\overline\rho}
\eea
is verified.\par
Since no confusion will arise, in both parabolic and hyperbolic cases, we denote in the following, for simplicity, the scaling parameter of the homogeneous $D$-module rep as ``$\lambda$" and the parameter of the inhomogeneous $D$-module rep as ``$\rho$".

\section{Conformal actions in $d=1$}

We are looking at first for conformally invariant actions depending on a single field $\varphi(t)$.
The Lagrangian has the form
\bea
{\cal L} &=& g(\varphi){\dot \varphi}^2+h(\varphi),
\eea
where $g(\varphi)$ is a (one-dimensional) metric and $h(\varphi)$ is a potential term. The conformal
invariance puts restrictions on both $g$ and $h$.\par
We present here the general results for the four types of conformal transformations (homogeneous/inhomogeneous and parabolic/hyperbolic) introduced in Sections {\bf 2} and {\bf 3}.\par
In the homogeneous parabolic case, the invariance under the $L_n$ transformations (\ref{hom})
requires solving the system of equations
\bea\label{homeq}
{\dot a}_n [(1+2\lambda)g+{\lambda}g_\varphi\varphi]&=&0,\nonumber\\
2\lambda g {\ddot a}_{n}\varphi +h_\varphi a_n + N^{(n)}_\varphi &=&0,\nonumber\\
\lambda h_\varphi\varphi {\dot a}_n + N^{(n)}_t &=& 0,
\eea
with $a_n$ given in (\ref{par}). The set of functions $N^{(n)}(\varphi, t)$ has to be determined; it reflects the arbitrariness of the invariance of the Lagrangian up to a total time-derivative.\par
The same system is derived in the homogeneous hyperbolic case with $a_n$ given in (\ref{hyp}). In the hyperbolic case we have the relation 
\bea \label{hyprel}
{\ddot a}_n &=& n^2\mu^2 a_n,
\eea
which is not present in the parabolic case.\par
Under the inhomogeneous transformations (\ref{inhom}) the system of equations 
\bea\label{inhomeq}
{\dot a}_n[g+\rho g_\varphi] &=&0,\nonumber\\
2\rho g{\ddot a}_n+h_\varphi {a}_n + N^{(n)}_\varphi &=& 0,\nonumber\\
\rho h_\varphi {\dot a}_n +N^{(n)}_t &=& 0
\eea
is derived for both parabolic  and hyperbolic cases; $a_n$ is given, respectively, by (\ref{par}) and
(\ref{hyp}). \par
Solving the above systems for all four cases is straightforward.\par
In the {\em Hom. par.} case, for instance, the first set of equations in (\ref{homeq}) gives for the metric the solution
$g = C_1\varphi^{-\frac{(1+2\lambda)}{\lambda}}$ ($C_1$ is a normalization constant). The third set of (\ref{homeq}) equations allows to write $N^{(n)}= -\lambda h_\varphi \varphi a_n+ M^{(n)}$, where $M^{(n)}(\varphi)$ are arbitrary functions of $\varphi$ which do not explicitly depend on the time coordinate $t$. By plugging this result into the second set of equations, together with the (\ref{par}) position
for $a_n$, we end up with the following system:
$-2\lambda(n+1)nt^{n-1}g\varphi -t^{n+1}[(1-\lambda) h_\varphi- \lambda h_{\varphi\varphi}\varphi]+M^{(n)}_\varphi =0$.\par
The vanishing of the term inside square brackets gives the solution $h= C_2\varphi^\frac{1}{\lambda}$ ($C_2$ is the normalization constant). The first term in the left hand side is vanishing for $n=0,-1$, while it can be reabsorbed by a suitable choice of $M^{(1)}(\varphi)$ for $n=1$. \par
Therefore, the (\ref{homeq}) system of equations cannot be nontrivially solved, simultaneously, for all $n\in {\mathbb{Z}}$, but at most for the $sl(2)$ subalgebra.\par
Deriving the solution for the three remaining cases proceeds along similar lines. In the two hyperbolic cases, the (\ref{hyprel}) relation for the $a_n$'s induces an extra term in the potential, proportional to the metric normalization constant $C_1$, which is not present in the parabolic cases.\par
The overall results can be summarized as follows. We obtain four $d=1$ conformal actions, invariant under
different realizations of the $sl(2)$ active transformations of the single bosonic field $\varphi(t)$. Their respective Lagrangians are given by\\

 {\em I - Homogeneous parabolic case}:
\bea\label{homparlag}
{\cal L}&=& C_1 \varphi^{-\frac{(1+2\lambda)}{\lambda}}{\dot\varphi}^2+C_2\varphi^\frac{1}{\lambda}.
\eea

 {\em II - Inhomogeneous parabolic case}:
\bea\label{inhparlag}
{\cal L}&=& C_1e^{-\frac{1}{\rho}\varphi}{\dot\varphi}^2+C_2e^{\frac{1}{\rho}\varphi}.
\eea

 {\em III - Homogeneous hyperbolic case}:
\bea\label{homhyplag}
{\cal L}&=& C_1 [\varphi^{-\frac{(1+2\lambda)}{\lambda}}{\dot\varphi}^2+
\mu^2{\lambda^2}\varphi^{-\frac{1}{\lambda}}]+
C_2\varphi^\frac{1}{\lambda}.
\eea

 {\em IV - Inhomogeneous hyperbolic case}:
\bea\label{inhhyplag}
{\cal L}&=& C_1e^{-\frac{1}{\rho}\varphi}[{\dot\varphi}^2+\mu^2{\rho}^2]+C_2e^{\frac{1}{\rho}\varphi}.
\eea

In order to have a dimensionless action ${\cal S}$ ($[{\cal S}]=0$), the scaling dimension of the Lagrangian
is $[{\cal L}]=1$, if we assign the time coordinate to have scaling dimension $-1$.
Taking into account $[\mu]=1$ and the relations (\ref{scalings}), we end up with the following dimensional analysis:\\
- in both homogeneous cases ({\em I} and {\em III}), $[C_1]=[C_2]=0$, provided that $[\varphi]={\lambda}$;\\
- in both inhomogeneous cases ({\em II} and {\em IV}), $[C_1]=-1-2s$, $[C_2]=1$, $[\varphi]=[\rho]=s$, with $s$ arbitrary.\par
The homogeneous parabolic case is the only one not containing dimensional constants in the conformal action (in the {\em Hom. hyp.} case the constant $\mu$ is present).\par
The $C_1, C_2$ constants are arbitrary. On the other hand, in the two hyperbolic cases, extra terms for the potential appear with respect to the parabolic cases. Their normalization constant ($C_1\mu^2$) is linked with the metric normalization constant. Since $\mu^2$ is positive, these potential terms have a ``wrong" sign and are not bounded below.\par
A consistent action with a correct, bounded below, potential is obtained by allowing $\mu$ to be a complex variable and performing the $\mu\mapsto i\mu$ Wick rotation. As recalled in Section {\bf 2}, this is tantamount to pass from the hyperbolic to the trigonometric version of the $D$-module representation. 
We find convenient to derive the trigonometric actions in terms of a complex field $\varphi$ (the real case can be recovered by setting its imaginary part equal to zero). A simple inspection shows that conformal, $sl(2)$-invariant actions based on the trigonometric $D$-module reps are given by\\

{\em i - Homogeneous trigonometric case}:
\bea\label{homtrilag}
{\cal L}&=& C_1 (|\varphi|^{-\frac{(1+2\lambda)}{\lambda}}{\dot\varphi}^\ast{\dot\varphi}-
\mu^2{\lambda^2}|\varphi|^{-\frac{1}{\lambda}})+
C_2|\varphi|^\frac{1}{\lambda};
\eea

{\em ii - Inhomogeneous trigonometric case}:
\bea\label{inhtrilag}
{\cal L}&=& C_1e^{-\frac{\varphi+\varphi^\ast}{2\rho}}({\dot\varphi}^\ast{\dot\varphi}-\mu^2{\rho}^2)+C_2e^{\frac{\varphi+\varphi^\ast}{2\rho}}.
\eea

In both cases the correct, bounded below potentials are obtained by choosing $C_1>0$ and $C_2\leq 0$.\par
Our results should be compared with the ones derived by Papadopoulos in \cite{pap}. In that paper only homogeneous transformations were considered. Furthermore, only constant metrics were discussed.
This amounts to set $\lambda= -\frac{1}{2}$ in the homogeneous parabolic Lagrangian (\ref{homparlag}) and
in the homogeneous hyperbolic Lagrangian (\ref{homhyplag}) (the results of \cite{pap} are recovered, as it should be, in these special cases).
These restrictions, however, can no longer be justified for the ${\cal N}$-extended superconformal actions with ${\cal N}\geq 4$. As already pointed out in Section {\bf 2}, in the ${\cal N}\geq 4$ cases, the parameter $\lambda$ becomes critical. It specifies under which of the inequivalent superconformal algebras the mechanical system is invariant. We postpone to Appendix {\bf C} (after the introduction of ${\cal N}=4$ superconformal actions in Section {\bf 9}) a discussion of the subtle issues concerning the relation of the inhomogenous versus homogeneous actions.


\section{Conformal actions in $d=2$}

It is instructive to extend the previous analysis to $d=2$ conformally invariant actions. 
In the Lagrangian framework and classical case, the infinite-dimensional conformal algebra
is ${\mathfrak{witt}}\oplus{\mathfrak{witt}}$, the direct sum of two copies of the Witt algebra ${\mathfrak{witt}}$.\par
Let $x_{1,2}$ be the coordinates of the plane. The $L_n^\pm$ generators of a $D$-module rep of
${\mathfrak{witt}}\oplus{\mathfrak{witt}}$ can be written in terms of the chiral/antichiral coordinates
$z_\pm = x_1\pm x_2$. The $L_n^\pm$ generators can be recovered from the $d=1$ $L_n$ generators 
introduced in Section {\bf 2} after replacing either $t$ or $\tau$ (in the respective cases) with $z_\pm$. The chiral/antichiral decomposition implies the vanishing of the commutators $[L_n^+,L_m^-]=0$ for any
$n,m\in {\mathbb Z}$.\par
The two-dimensional conformal actions have a Lagrangian of the form
\bea
{\cal L}&=& g\varphi_+\varphi_-+h,
\eea
where the $\pm$ suffix denotes the partial derivative with respect to $z_\pm$.\\
Looking for conformal invariance under the assumption that homogeneous/homogeneous or inhomogeneous/inhomogeneous active D-module transformations of $\varphi (z_\pm)$ apply on both  chiral/antichiral sectors, we are led to the following results. Contrary to the $d=1$ case, the parabolic and hyperbolic $D$-module reps produce the same output for the Lagrangians, while the actions are invariant under
the whole infinite set of $L_n^\pm$ generators.\par
The two surviving cases correspond to the Homogeneous or the Inhomogeneous transformations, respectively.\par
In the Homogeneous case $g,h$ are restricted so that the conformally invariant Lagrangian is
\bea\label{d2hom}
{\cal L}&=& \frac{C_1}{\varphi^2}\varphi_+\varphi_- + C_2 \varphi^{\frac{1}{\lambda}},
\eea
with $C_1,C_2$ arbitrary constants.\par 
The resulting action is invariant under the
$\delta_n^\pm (\varphi) = - z_\pm^{n+1}\varphi_\pm -\lambda(n+1) z_\pm^n\varphi$ transformations.\par
In the Inhomogeneous case we recover the Liouville action. The metric needs to be a constant, while the potential is the exponential Liouville potential. We have
\bea\label{d2liou}
{\cal L}&=& C_1\phi_+\phi_- +C_2e^{\frac{\phi}{\rho}}.
\eea
The corresponding action is invariant under the $\delta_n^\pm (\phi) = - z_\pm^{n+1}\phi_\pm -\rho (n+1) z_\pm^n$ transformations.\par
By requiring the action being dimensionless, and assuming $[z_\pm]=-1$, we obtain the Lagrangian scaling dimension $[{\cal L}]=2$. \par
In the Homogeneous case the scaling dimensions are fixed to be
\bea
&
[\varphi]=2\lambda,\quad [C_1]=[C_2]=[\lambda]=0
&
\eea (therefore, no dimensional parameter is present in the theory).\\
In the Inhomogenous (Liouville) case, for an arbitrary value $s$, the scaling dimensions are
\bea
&[\phi]=[\rho]=s,\quad [C_1]=-2s,\quad [C_2]= 2.
&
\eea

In this two-dimensional case the homogeneous action (\ref{d2hom}) is recovered from the Liouville action
(\ref{d2liou}) through the non-linear field redefinition $\varphi=e^\phi$ and by performing the identification
$\lambda=\rho$ (as discussed in Appendix {\bf C} the latter identification is not possible for systems
that, unlike (\ref{d2hom}) and (\ref{d2liou}), possess a critical value of the scaling dimension $\lambda$).\par
One should note that the classical Liouville action is invariant under two separate copies of the
{\em centerless} Virasoro algebra.  Even in this case, on the other hand, the associated Noether charges, endowed with a Poisson brackets structure, necessarily close the centrally extended version of the algebra, the full $Vir\oplus Vir$ algebra.
It is a consequence of a non-equivariant moment map applied to the Liouville theory (see \cite{topnlin} for details).

\section{On superconformal algebras}

We discuss here two types of superconformal algebras, the supersymmetric extensions of the $d=1$ conformal algebra $sl(2)$ and the supersymmetric extensions of the Virasoro algebra.\par
The finite one-dimensional superconformal algebras belong to the simple Lie superalgebras classified in \cite{{kac},{nah},{dictionary}} and satisfy special properties. A $d=1$ superconformal algebra ${\cal G}$
admits a grading \cite{toptianjin}
${\cal G} = {\cal G}_{-1}\oplus {\cal G}_{-\frac{1}{2}}\oplus
{\cal G}_0\oplus {\cal G}_{\frac{1}{2}}\oplus {\cal G}_{1}$. Its even sector ${\cal G}_{even}={\cal G}_0\oplus {\cal G}_{-1}\oplus {\cal G}_1$ is isomorphic to $sl(2)\oplus R$, where the subalgebra $R$ is known as $R$-symmetry. The odd sector (${\cal G}_{\frac{1}{2}}\oplus {\cal G}_{-\frac{1}{2}}$) is spanned by $2{\cal N}$ generators (${\cal N}$ is the number of extended supersymmetries).\\
At fixed ${\cal N}$ the positive sector ${\cal G}_{>0}$ is isomorphic to the $d=1$ superPoincar\'e algebra (the algebra of the ${\cal N}$-extended supersymmetric quantum mechanics \cite{wit}).\par
If we denote, see ({\ref{sl2}), the $sl(2)$ generators as $D,H,K$, we have that ${\cal G}_1$ (${\cal G}_{-1}$) is spanned by the positive
(negative) root $H$ ($K$), while ${\cal G}_0=D{\mathbb C}\oplus R$.\par
We are especially interested in the ${\cal N}=1,2,4,8$ extensions. The corresponding list of $d=1$ superconformal algebras is given by
$osp(1|2)$ for ${\cal N}=1$ and $sl(2|1)$ for ${\cal N}=2$. For ${\cal N}=4$ we have the exceptional superalgebras $D(2,1;\alpha)$, depending on the complex parameter $\alpha \neq 0,-1$ and $A(1,1)=
sl(2|2)/{\mathbb Z}$ (it can be recovered for $\alpha=0,-1$). Four distinct simple Lie superalgebras exist for ${\cal N}=8$: $A(3,1)$, $D(4,1)$, $D(2,2)$ and the exceptional superalgebra $F(4)$.\par 
The exceptional superalgebras $D(2,1;\alpha)$, $D(2,1;\alpha')$ are isomorphic iff $\alpha'$ belongs to an $S_3$-group orbit generated by the moves $\alpha\mapsto\frac{1}{\alpha}$ and $\alpha\mapsto -(1+\alpha)$, i.e. if $\alpha'$ takes one of the six values
\bea\label{S3orbit}
&\alpha, \quad\frac{1}{\alpha},\quad -(1+\alpha),\quad -\frac{1}{(1+\alpha)},\quad -\frac{(1+\alpha)}{\alpha},\quad -\frac{\alpha}{(1+\alpha)}.&
\eea
The (homogeneous and parabolic) $D$-module reps of the above $d=1$ superconformal algebras have been constructed in \cite{kuto} (the ${\cal N}=1,2,4$ cases and one ${\cal N}=8$ example) and \cite{khto} (the remaining ${\cal N}=8$ cases). The construction relies upon the
classification, presented in \cite{pato} and \cite{kuroto}, of the $d=1$ superPoincar\'e (the ${\cal G}_{>0}$ subalgebra) $D$-module reps.
\par
Concerning the $d=1$ superPoincar\'e $D$-module reps for ${\cal N}=1,2,4,8$, the results can be summarized as follows. The differential operators act on ${\cal N}$ bosonic and ${\cal N}$ fermionic fields (the supermultiplet). For any $k=0, 1,\ldots, {\cal N}$, we have $k$ fields with scaling dimension
$\lambda$ (they are known as the ``propagating bosons"), ${\cal N}$ fields (the fermions) with scaling dimension $\lambda+\frac{1}{2}$
and the remaining ${\cal N}-k$ fields (the so-called ``auxiliary bosons") with scaling dimension $\lambda+1$. Both a supermultiplet and the associated $d=1$ superPoincar\'e $D$-module rep will be denoted with the symbol ``$(k,{\cal N},{\cal N}-k)_\lambda$".\par
The extension to a $d=1$ superconformal algebra $D$-module rep requires introducing (in compatible way, so that to close the (anti)commutation relations) the extra differential operators associated to the ${\cal G}_{\leq 0}$ generators.\par
The \cite{kuto} and \cite{khto} results can be summarized as follows:
\\
{\em i}) for ${\cal N}=1,2$ and any value of the scaling dimension $\lambda$ (no criticality), the $(k,{\cal N}, {\cal N}-k)_\lambda$ supermultiplet induces a $D$-module rep for $osp(1|2)$ and $sl(2|1)$, respectively;\\
{\em ii}) for ${\cal N}=4$ the $(k,{\cal N}, {\cal N}-k)_\lambda$ supermultiplet induces a $D$-module rep for the $D(2,1;\alpha)$ superalgebra with the identification 
\bea\label{alphalambda}
\alpha &=& (2-k)\lambda
\eea 
(since $\alpha$, up to the (\ref{S3orbit}) relations, parametrizes inequivalent superalgebras, we already encounter here the criticality of the scaling dimension);\\
{\em iii}) for ${\cal N}=8$ the $(k,{\cal N}, {\cal N}-k)_\lambda$ supermultiplet induces a $D$-module rep for a superconformal algebra
only for $k\neq 4$ and at the critical scaling dimensions 
\bea
&\lambda\equiv \lambda_k= \frac{1}{k-4};&
\eea
 the given superalgebras are $D(4,1)$for $k=0,8$, $F(4)$ for $k=1,7$, $A(3,1)$ for $k=2,6$ and $D(2,2)$ for $k=3,5$.\par
The $D$-module reps for the ${\cal N}=4$ $d=1$ superconformal algebra $A(1,1)$ (recovered from the $\alpha=0,-1$ values) are, in particular, obtained at the critical values 
\bea
\lambda =0 \quad& {\mbox{and}} &\quad\lambda =\frac{1}{k-2}\quad (k\neq 2)
\eea 
for, respectively, the supermultiplets
\bea
(k,4,4-k)_{\lambda =0}\quad \forall k=0,1,2,3,4 &{\mbox{and}}& (k,4,4-k)_{\lambda=\frac{1}{k-2}},\quad k\neq 2.
\eea
The singular transformation, discussed in Section {\bf 2}, which relates the parabolic and hyperbolic $D$-module reps of $sl(2)$
(producing, in particular, the (\ref{scalingequality}) equality between the respective scaling dimensions) is applicable in all supersymmetric cases. As a consequence, homogeneous hyperbolic $D$-module reps, with the same criticality properties of the corresponding
parabolic cases, are immediately obtained for all the above listed superalgebras.\par
For what concerns the supersymmetric extensions of the Virasoro algebras, it is known \cite{gls} that non-trivial central charges can only exist up
to ${\cal N}=4$. Since in the following we are dealing with $D$-module reps, here we only need to consider the {\em centerless} ($c=0$) ${\cal N}=1,2,4$ superVirasoro algebras which generalize the Witt algebra. \par
The centerless ${\cal N}=4$ superVirasoro algebra is spanned by the even generators
$L_n$, $J^i_n$ and by the odd generators $Q^I_r$, where $I=0,1,2,3$ and
$i=1,2,3$. The centerless ${\cal N}=1,2$ superVirasoro algebras are its subalgebras, obtained by restricting $I=0,1$ and $i=1$ for ${\cal N}=2$ and $I=0$ for ${\cal N}=1$ (the latter case includes only the $L_n, Q^0_r$ generators). Two variants of the superalgebras exist \cite{gsw}, the Ramond (R) and the Neveu-Schwarz (NS) versions. In both cases $n$ is an integer ($n\in {\mathbb Z}$); in the Ramond version $r$ is also an integer ($r\in {\mathbb Z}$), while in the Neveu-Schwarz version $r$ is a half-integer number ($r\in\frac{1}{2}+{\mathbb Z}$).\par
The (anti)commutators of the centerless ${\cal N}=4$ superVirasoro algebra are
explicitly given by
\bea\label{svir}
\relax [L_m, L_n]&=& (m-n) L_{m+n},\nonumber\\
\relax [L_n, Q^I_r] &=& (\frac{n}{2}-r)Q^I_{n+r},\nonumber\\
\relax [ L_n, J^i_m]&=& -mJ^i_{n+m},\nonumber\\
\{Q^0_r,Q^0_s\}&=& 2L_{r+s},\nonumber\\
\{Q^0_r,Q^i_s\}&=& 2(r-s)J^i_{r+s},\nonumber\\
\{Q^i_r,Q^j_s\}&=& 2\delta^{ij}L_{r+s}+2\epsilon^{ijk}(r-s)J^k_{r+s},\nonumber\\
\relax [ Q^0_r, J^i_n]&=& \frac{1}{2}Q^i_{n+r},\nonumber\\
\relax [ Q^i_r, J^j_n]&=& -\frac{1}{2}\delta^{ij}Q^0_{n+r}-\frac{1}{2}\epsilon^{ijk}Q^k_{n+r},\nonumber\\
\relax [J^i_n,J^j_m]&=& -\epsilon^{ijk} J^k_{n+m}.
\eea

The finite $d=1$ ${\cal N}=4$ superconformal algebra $A(1,1)$ is recovered as a subalgebra. \par
In the Ramond version the $A(1,1)$ generators are
$L_0, L_{\pm 2}, Q^I_{\pm 1}, J^i_0$;\par
 in the Neveu-Schwarz version they are
$L_{\pm 1}, L_0, Q^I_{\pm \frac{1}{2}}, J^i_0$.\par
The $osp(1|2)$ subalgebra is given by the generators \par
$L_0, L_{\pm 1}, Q^0_{\pm \frac{1}{2}}$ (NS) or
$L_0, L_{\pm 2}, Q^0_{\pm 1} $ (R).\par
The $sl(2|1)$ subalgebra is given by the generators \par
$L_0, L_{\pm 1}, Q^0_{\pm \frac{1}{2}}, Q^1_{\pm \frac{1}{2}}, J^1_0$ (NS) or
$L_0, L_{\pm 2}, Q^0_{\pm 1}, Q^1_{\pm 1}, J^1_0$ (R).\par

An extra centerless superVirasoro case which does not fit into the above scheme and contains an extra set of odd generators ($W_r$) is given by the ${\cal N}=3$
extension. The even generators of the ${\cal N}=3$ centerless superVirasoro are $L_n$ and $J^i_n$,
while the odd generators are $Q^i_r$ and $W_r$ ($i=1,2,3$).
The (anti)commutators are explicitly given by
\bea\label{n3}
\relax [L_m, L_n]&=& (m-n) L_{m+n},\nonumber\\
\relax [L_n, Q^i_r] &=& (\frac{n}{2}-r)Q^i_{n+r},\nonumber\\
\relax [ L_n, J^i_m]&=& -mJ^i_{n+m},\nonumber\\
\relax [L_n, W_r] &=& -(\frac{n}{2}+r)W_{n+r},\nonumber\\
\{Q^i_r,Q^j_s\}&=& 2\delta^{ij}L_{r+s}+2\epsilon^{ijk}(r-s)J^k_{r+s},\nonumber\\
\relax [Q^i_r, J^j_n]&=& -\frac{1}{2}\epsilon^{ijk}Q^k_{n+r}-\frac{n}{2}\delta^{ij}W_{n+r},\nonumber\\
\relax \{ Q^i_r, W_s\}&=& 2J^i_{r+s},\nonumber\\
\relax [J^i_n,J^j_m]&=& -\frac{1}{2}\epsilon^{ijk} J^k_{n+m},\nonumber\\
\relax [J^i_n, W_r] &=& 0,\nonumber\\
\relax \{W_r,W_s\}&=& 0.
\eea
 The finite subalgebra consisting of the twelve generators $L_0, L_{\pm 1}, J^i_0, Q^i_{\pm \frac{1}{2}}$
(please note the absence of the $W_r$'s generators)
is the $d=1$  ${\cal N}=3$ superconformal algebra $B(1,1)=osp(3|2)$.\par
 In \cite{khto} a $D$-module rep for $B(1,1)$ 
was constructed. It acts on the $(1,3,3,1)$ supermultiplet which contains one bosonic field
of scaling dimension $\lambda$, three fermionic fields
of scaling dimension $\lambda+\frac{1}{2}$, three bosonic fields of scaling dimension $\lambda+1$ and one fermionic field of scaling dimension $\lambda+\frac{3}{2}$. This $D$-module rep (existing for an arbitrary $\lambda$) is non-critical.

\section{New results for superconformal $D$-module reps}

In this Section we concentrate all new results concerning $D$-module reps
of superconformal algebras. Our analysis heavily used algebraic computations with {\em Mathematica}.\par
Two classes of results are presented. At first we extend the construction presented in \cite{kuto} and \cite{khto} of the {\em homogeneous} $D$-module reps to the
case of the {\em inhomogeneous} $D$-module reps of the $d=1$ finite superconformal algebras (the ones we introduced in Section {\bf 6}). Next, we extend the
\cite{kuto} and \cite{khto} results to the case of $D$-module reps
(both {\em homogeneous} and {\em inhomogeneous}) of the centerless ${\cal N}=1,2,3,4$ superVirasoro algebras.\par
The explicit presentation of these $D$-module reps is given in Appendix {\bf B}.
\par
As discussed in Appendix {\bf A}, the new class of inhomogeneous $D$-module reps  are obtained for $\lambda=0$ and $\rho\neq 0$.\par
Since the presence of at least a propagating boson is required to construct the inhomogeneous term, the inhomogeneous supermultiplets $(k, {\cal N}, {\cal N}-k)_{{\lambda=0,\rho\neq 0}}$ can only exist for $k\geq 1$.\par
 The list of inhomogeneous $D$-module reps for the finite $d=1$ superconformal algebras of Section {\bf 6}
is given by
\bea\label{inhomrepsusy}
{\cal N}=1 &:&  osp(1|2) \quad {\mbox{with}} \quad (1,1)_{0,\rho},\nonumber\\
{\cal N}=2 &:&  sl(2|1) \quad ~~{\mbox{with}} \quad (1,2,1)_{0,\rho}, \quad (2,2)_{0,\rho}, \nonumber\\
{\cal N}=3 &:& B(1,1) \quad ~~{\mbox{with}}\quad (1,3,3,1)_{0,\rho},\nonumber\\
{\cal N}=4 &:&  A(1,1) \quad~~ {\mbox{with}} \quad (1,4,3)_{0,\rho}, \quad (2,4,2)_{0,\rho},\quad (3,4,1)_{0,\rho}, \quad (4,4,0)_{0,\rho},\nonumber\\
{\cal N}=8 &:& {\mbox{none}}
\eea
(the last result is a consequence of the fact that, for ${\cal N}=8$, the $d=1$ finite superconformal algebras 
have critical scalings $\lambda\neq 0$). \par 
Concerning the centerless superVirasoro algebras, the homogeneous supermultiplets are encountered for
\bea\label{homrepsvir}
{\cal N}=1 &{\mbox{SVir}}:&   (k,1,1-k)_{\lambda},\quad k=0,1\quad {\mbox{with}}\quad\lambda\quad {\mbox{arbitrary}},\nonumber\\
{\cal N}=2 &{\mbox{SVir}}:&  (k,2,2-k)_\lambda, \quad k=0,1,2 \quad {\mbox{with}}\quad \lambda\quad {\mbox{arbitrary}}, \nonumber\\
{\cal N}=3 &{\mbox{SVir}}:& (1,3,3,1)_\lambda, \quad\quad {\mbox{with}}\quad \lambda\quad{\mbox{arbitrary}},\nonumber\\
{\cal N}=4 &{\mbox{SVir}}:&  (k,4,4-k)_\lambda, \quad k=0,1,2,3,4\quad {\mbox{with}} \quad \lambda=0 \quad {\mbox{or}}\quad  \lambda = \frac{1}{k-2} (k\neq 2).\nonumber\\&&
\eea

The inhomogeneous $D$-module reps of the centerless superVirasoro algebras are only encountered for ${\cal N}=1,2, 3$ but not for ${\cal N}=4$:
\bea\label{inhomrepsvir}
{\cal N}=1 &{\mbox{SVir}}:&   (1,1)_{0,\rho},\nonumber\\
{\cal N}=2 &{\mbox{SVir}}:&  (2,2,0)_{0,\rho}\quad {\mbox{and}}\quad (1,2,1)_{0,\rho},\nonumber\\
{\cal N}=3 &{\mbox{SVir}}:& (1,3,3,1)_{0,\rho},\nonumber\\
{\cal N}=4 &{\mbox{SVir}}:& {\mbox{none}}.
\eea

It is instructive to show the reason for 
the absence of the inhomogeneous $D$-module reps for the centerless ${\cal N}=4$ superVirasoro. It is due to the fact that, in particular, the closure of the algebra requires the commutators $[J^3_n, Q^3_r]$ to be proportional to the $Q^4_{n+r}$ generators. Let's take, as an example, the $(4,4,0)_{\lambda,\rho}$ supermultiplet. We are led to a system of equations to be solved:
\bea
2\lambda-\frac{1}{2}&=& A,\nonumber\\
(\frac{1}{2}-2\lambda)\partial_t+\lambda(-n+r(1-4\lambda))&=& A(-\partial_t-2(n+r)\lambda),\nonumber\\  
-(n+r(4\lambda-1))\rho &=& -2A(n+r)\rho,
\eea
where $A$ is a proportionality constant.\par
To solve the system for all $n,r$, either one has to set $\lambda=\frac{1}{2}$ and $\rho$ arbitrary (which is equivalent to the homogeneous representation  $(4,4,0)_\frac{1}{2}$) or $\lambda=\rho=0$.\par
By restricting the conditions to $n=0$ and $r=\pm {\overline r}$ (the case of the $A(1,1)$ subalgebra), the system is solved for arbitrary values of $\lambda$ and $\rho$.\par
The inspection of the consistency  conditions induced by all (anti)commutators leads to the results that we have presented in this Section.\par
It is worth pointing out, as a last comment, that the inhomogeneous $D$-module reps discussed here consist of a different and inequivalent class of linear transformations with respect to the inhomogenous 
$D$-module reps discussed in \cite{kuto} and \cite{khto}. There, the $sl(2)$ generators act homogeneously
and the representations are only obtained at the critical value $\lambda=-1$.

\section{Some examples of ${\cal N}=1,2$ superconformal actions in $d=1,2$ dimensions}

We illustrate here an application of supersymmetry with the construction of some ${\cal N}=1,2$ superconformal actions in $d=1,2$ dimensions.\par
In $d=1$ we obtain the following $osp(1|2)$-invariant actions for the ${\cal N}=1$ supermultiplet $(1,1)$
(a single boson $\varphi$ and a single fermion $\psi$):\\
 
{\em I} - for the {\em homogeneous parabolic} case the Lagrangian is
\bea\label{n1hompar}
{\cal L}&=& C\varphi^{-\frac{1+2\lambda}{\lambda}}({\dot \varphi}^2+\psi{\dot\psi}),
\eea

with dimensions $[\varphi]=\lambda$, $[\psi]=\lambda+\frac{1}{2}$, $[C]=[\lambda]=0$;\par
{\em II} - for the {\em inhomogeneous parabolic} case the Lagrangian is
\bea\label{n1inhpar}
{\cal L}&=& Ce^{-\frac{\varphi}{\rho}}({\dot \varphi}^2+\psi{\dot\psi}),
\eea

with dimensions $[\varphi]=[\rho]=s$, $[\psi]=s+\frac{1}{2}$, $[C]=-1-2s$;\par
{\em III} - for the {\em homogeneous hyperbolic} case the Lagrangian is
\bea\label{n1homhyp}
{\cal L}&=& C[\varphi^{-\frac{1+2\lambda}{\lambda}}({\dot \varphi}^2+\mu\psi{\dot\psi})
+\mu^2\lambda^2\varphi^{-\frac{1}{\lambda}}],
\eea

with dimensions $[\varphi]=[\psi]=\lambda$, $[\mu]=1$, $[C]=[\lambda]=0$;\par
{\em IV} - for the {\em inhomogeneous hyperbolic} case the Lagrangian is
\bea\label{n1inhhyp}
{\cal L}&=& Ce^{-\frac{\varphi}{\rho}}({\dot \varphi}^2+\mu\psi{\dot\psi}+\mu^2\rho^2),
\eea

with dimensions $[\varphi]=[\psi]=[\rho]=s$, $[\mu]=1$, $[C]=-1-2s$.\par
We note a similarity and a difference with respect to the purely bosonic results. Just like the bosonic actions,
the hyperbolic cases present a potential term, proportional to $C_1\mu^2$, which is absent in the parabolic cases. On the other hand, the potential terms proportional to $C_2$ and appearing in (\ref{homparlag}-\ref{inhhyplag}) are now excluded due to the supersymmetry constraint.\par

In $d=2$, for a single boson $\varphi$ and a single fermion $\psi$, we obtain ${\cal N}=1$ chiral (antichiral) actions, invariant under a {\em single} ({\em chiral/antichiral}) {\em copy} of the centerless superVirasoro algebra. The Lagrangians are given by\par
\bea\label{chiralhomd2}
{\cal L}^\pm &=& \frac{C}{\varphi^2}(\varphi_+\varphi_-+\psi\psi_{\mp}),
\eea
for the homogeneous case and
\bea\label{chiralinhd2}
{\cal L}^{\pm}&=& C(\varphi_+\varphi_-+\psi\psi_\mp)
\eea
(the constant kinetic term) for the inhomogeneous case.\par
It should be pointed out that, in order to get the superLiouville extension, a second fermion needs to be added, see \cite{sliou} and \cite{topsliou}.\par
As an ${\cal N}=2$ example in $d=1$ we present the $sl(2|1)$-invariant actions for the supermultiplet
$(1,2,1)$ (a propagating boson $\varphi$, two fermions $\psi_1,\psi_2$ and an auxiliary bosonic field $g$).
The Lagrangians are:\\

{\em I} - for the {\em homogeneous parabolic} case
\bea\label{n2hompar}
{\cal L}&=& A({\dot\varphi}^2+\psi_i{\dot\psi}_i+g^2) -\frac{1}{2}A_\varphi\epsilon^{ij}\psi_i\psi_jg,
\quad \quad A= C\varphi^{-\frac{1+2\lambda}{\lambda}};
\eea

{\em II} - for the {\em inhomogeneous parabolic} case
\bea\label{n2inhpar}
{\cal L}&=& A({\dot\varphi}^2+\psi_i{\dot\psi}_i+g^2) -\frac{1}{2}A_\varphi\epsilon^{ij}\psi_i\psi_jg,
\quad \quad A= Ce^{-\frac{\varphi}{\rho}};
\eea

{\em III} - for the {\em homogeneous hyperbolic} case
\bea\label{n2homhyp}
{\cal L}&=& A({\dot\varphi}^2+\mu\psi_i{\dot\psi}_i+\mu^2g^2) -\frac{1}{2}\mu^2A_\varphi\epsilon^{ij}\psi_i\psi_jg+\mu^2\lambda^2A\varphi^2,
\quad \quad A= C\varphi^{-\frac{1+2\lambda}{\lambda}};
\eea

{\em IV} - for the {\em inhomogeneous hyperbolic} case
\bea\label{n2inhhyp}
{\cal L}&=& A({\dot\varphi}^2+\mu\psi_i{\dot\psi}_i+\mu^2g^2) -\frac{1}{2}\mu^2A_\varphi\epsilon^{ij}\psi_i\psi_jg+\mu^2\rho^2A,
\quad \quad A= Ce^{-\frac{\varphi}{\rho}}.
\eea

\section{On ${\cal N}=4~
d=1$ superconformal actions with exceptional $D(2,1;\alpha)$ invariance}

The supermultiplet $(1,4,3)_\lambda$ consists of a single propagating boson $\varphi$, four fermions $ \psi_0, \psi_i$ and three auxiliary bosons $ g_i$ (here $i=1,2,3$; in the formulas below we also use the index 
$I=0,1,2,3$).
For this supermultiplet, see formula (\ref{alphalambda}), we have $\alpha=\lambda$. We list here its superconformally invariant actions. \\
For homogeneous transformations the Lagrangians of the $D(2,1;\alpha)$-invariant actions are,\\
in the {\em homogeneous parabolic} case,
\bea\label{n4hompar}
{\cal L}&=& A({\dot\varphi}^2+\psi_I{\dot\psi}_I+g_i^2) + A_\varphi(\psi_0\psi_ig_i +\frac{1}{2}\epsilon^{ijk}
\psi_i\psi_jg_k)+\frac{1}{6}A_{\varphi\varphi}\epsilon^{ijk}\psi_0\psi_i\psi_j\psi_k,\nonumber\\
 &{\mbox{with}} & A= C\varphi^{-\frac{1+2\alpha}{\alpha}}
\eea
and, in the {\em homogeneous hyperbolic} case,
\bea\label{n4homhyp}
{\cal L}&=& A({\dot\varphi}^2+\mu\psi_I{\dot\psi}_I+\mu^2g_i^2) + \mu^2A_\varphi(\psi_0\psi_ig_i +\frac{1}{2}\epsilon^{ijk}
\psi_i\psi_jg_k)+\nonumber\\
&& \frac{1}{6}\mu^2A_{\varphi\varphi}\epsilon^{ijk}\psi_0\psi_i\psi_j\psi_k
+\mu^2\alpha^2A{\varphi}^2,\nonumber\\
  &{\mbox{with}} & A= C\varphi^{-\frac{1+2\alpha}{\alpha}}. 
\eea
For inhomogeneous transformations, the requirement that $\rho\neq 0$ with $\lambda=0$ implies that
the superconformal actions are only invariant under the $A(1,1)$ superalgebra.\\
For inhomogeneous transformations the Lagrangians of the $A(1,1)$-invariant actions are,\\
in the {\em inhomogeneous parabolic} case,
\bea\label{n4inhpar}
{\cal L}&=& A({\dot\varphi}^2+\psi_I{\dot\psi}_I+g_i^2) + A_\varphi(\psi_0\psi_ig_i +\frac{1}{2}\epsilon^{ijk}
\psi_i\psi_jg_k)+\frac{1}{6}A_{\varphi\varphi}\epsilon^{ijk}\psi_0\psi_i\psi_j\psi_k,\nonumber\\
  &{\mbox{with}} & A= Ce^{-\frac{\varphi}{\rho}}
\eea
and,
in the {\em inhomogeneous hyperbolic} case, 
\bea\label{n4inhhyp}
{\cal L}&=& A({\dot\varphi}^2+\mu\psi_I{\dot\psi}_I+\mu^2g_i^2) +\mu^2 A_\varphi(\psi_0\psi_ig_i +\frac{1}{2}\epsilon^{ijk}
\psi_i\psi_jg_k)+\nonumber\\
&&\frac{1}{6}\mu^2A_{\varphi\varphi}\epsilon^{ijk}\psi_0\psi_i\psi_j\psi_k +\mu^2\rho^2A,\nonumber\\
  &{\mbox{with}} & A= Ce^{-\frac{\varphi}{\rho}}. 
\eea

\section{Conclusions}

We summarize the results of the paper. For what concerns representations, besides the results on the bosonic $sl(2)$ and Witt algebras, we introduced the new class of linear inhomogeneous $D$-module reps for the finite simple Lie superalgebras $osp(1|2)$, $sl(2|1)$, $B(1,1)$ and $A(1,1)$. 
These new reps, at the scaling dimension $\lambda=0$, depend on the parameter $\rho$ which measures
the inhomogeneity.\par $D$-module reps have also been constructed for the centerless superVirasoro algebras: homogeneous reps for the ${\cal N}=1,2,3,4$ extensions and inhomogeneous reps for the ${\cal N}=1,2,3$
extensions. They are based on the $(k, {\cal N}, {\cal N}-k)$ supermultiplets (for ${\cal N}=1,2,4$) and on
the $(1,3,3,1)$ supermultiplet (for ${\cal N}=3$).\par
We pointed out that, for both homogeneous and inhomogeneous reps,  two variants of the $D$-module reps
can be presented: parabolic and hyperbolic/trigonometric. They are mutually related by a singular transformation.\par
We ended up with four different types of active (super)conformal transformations ({\em hom. par.}, {\em inhom. par.}, {\em hom. hyp.} and {\em inhom. hyp.}) that can be used, in the Lagrangian setting, to construct (super)conformal actions in $d=1$ and $d=2$ dimensions. \par
For systems with ${\cal N}=0,1,2,4$ we presented the four types of $d=1$ actions invariant under their respective finite (super)conformal algebra.\par
In $d=2$ a non-linear field redefinition relates the two classes of homogeneous and inhomogeneous conformal actions exemplified by (\ref{d2hom}) and by the Liouville action (\ref{d2liou}). \par
The $d=1$ (super)conformal actions contain no dimensional constants only in the homogeneous parabolic case. This is the class of theories discussed in the \cite{fil3} review.  New classes of superconformally invariant theories can therefore be constructed from the three remaining types of transformations.\par 
It is rather straightforward to extend the results here presented to more complicated cases. For ${\cal N}=4$, for instance, one can investigate multi-particle systems by applying our construction to a certain number of supermultiplets in interactions. The provision is that the supermultiplets should carry a representation of the same superconformal
algebra (either $A(1,1)$ or $D(2,1;\alpha)$ for a fixed $\alpha$).\par
One of the possible interesting applications of our work  concerns the investigations on the $CFT(1)/AdS(2)$
correspondence (see \cite{sen} and \cite{cjps}). Our results shed a new light on the left side (the conformal side) of the correspondence.\par
A very promising field of investigation concerns the extension to non-relativistic conformal Galilei or conformal Newton-Hooke systems (see \cite{{gama},{aggm}}). Recently, a lot of activity in constructing models for this kind of theories has been motivated by the $CFT/AdS$ correspondence applied to non-relativistic systems like the ones appearing in condensed matter
(see, e.g., \cite{{har},{sac}}). Unlike the $(1+0)$-dimensional theories considered here, these conformal systems live
in $(1+d)$-dimension, $d$ being the number of space coordinates. A recent paper \cite{akt} proved how the (homogeneous parabolic) ${\cal N}=2$ superconformal $D$-module reps in $(1+0)$ can be enlarged to
induce ${\cal N}=2$ ${\ell}$-conformal Galilei superalgebras in $(1+d)$ dimensions. It is tempting to extend the new class of one-dimensional superconformal $D$-module reps discussed here to the $(1+d)$-dimensional case.

\par
{~}\par
\renewcommand{\theequation}{A.\arabic{equation}}
\setcounter{equation}{0}
 
{\Large{\bf Appendix A: On $D$-module reps and the interpolation of {\em Hom} and {\em Inhom} conformal actions}}\par
~\par
In principle one can ``mix" the homogeneous and inhomogeneous $D$-module reps of the Witt algebra
by allowing the couple of parameters
$(\lambda,\rho)$ being simultaneously non-vanishing. In the parabolic case, for instance, the general Witt algebra transformations applied on the field $\varphi$ are 
\bea\label{mixed}
L^{par}_m(\varphi)&=& -t^{m+1}{\dot\varphi}-\lambda(m+\gamma)t^m\varphi-\rho (m+\beta)t^m.
\eea
$L^{par}_{-1}$ is proportional to the Hamiltonian if we set $\gamma=1$  and $\beta=1$.\par
For $\lambda\neq 0$ we can write
 \bea
L^{par}_m(\varphi)&=& -t^{m+1}{\dot\varphi}-\lambda(m+1)t^m(\varphi+\frac{\rho}{\lambda}),
\eea
so that the action of the homogeneous transformation with scaling dimension $\lambda\neq 0$ is recovered for the shifted field $\overline\varphi= \varphi+\frac{\rho}{\lambda}$.
Therefore the $(\lambda,\rho)$ transformations with $\lambda\neq 0$ are equivalent to the pure homogeneous transformations with scaling parameter $\lambda$ and $\rho =0$. 
The same is true in the hyperbolic case.\par 
(\ref{mixed}) fails to interpolate the two cases, leaving us with the two inequivalent classes of\par
{\em i}) 
$(\lambda,0)$ homogeneous and\par
{\em ii}) $(0,\rho)$ inhomogeneous transformations.\par
The (\ref{mixed}) transformations, on the other hand, are useful to interpolate the conformally invariant actions. An $sl(2)$-invariant action (for $m=\pm 1, 0$) based on (\ref{mixed}) is given by the Lagrangian 
\bea\label{interpolag}
{\cal L} &=& K_1(\lambda\varphi +\rho)^{-\frac{1+2\lambda}{\lambda}}{\dot\varphi}^2+K_2(\lambda\varphi+\rho)^{\frac{1}{\lambda}},
\eea
with $K_1,K_2$ arbitrary constants. \par
The homogeneous Lagrangian (\ref{homparlag}) is recovered for $\rho =0$. \par
The inhomogeneous Lagrangian (\ref{inhparlag}) is recovered
in the $\lambda\rightarrow 0$ limit by suitably rescaling the constants $K_1,K_2$. This is accomplished by expressing the  (\ref{interpolag})  Lagrangian as
$$
{\cal L} = K_1\rho^{-\frac{1+2\lambda}{\lambda}}\left(1+\frac{\lambda\varphi}{\rho}\right)^{-\frac{1}{\lambda}-2}{\dot\varphi}^2+K_2\rho^{\frac{1}{\lambda}}\left(1+\frac{\lambda\varphi}{\rho}\right)^{\frac{1}{\lambda}}
$$
and by taking $K_1=C_1\rho^{\frac{1+2\lambda}{\lambda}}$ and $K_2=C_2\rho^{-\frac{1}{\lambda}}$.\par 
The possibility offered by the interpolation allows simplifying the constructions of the homogeneous and inhomogeneous conformally invariant actions, since both actions can be derived at one stroke.\par
The extension of the properties here discussed to the supersymmetric cases is immediate.

{~}\par
\renewcommand{\theequation}{B.\arabic{equation}}
\setcounter{equation}{0}
 
{\Large{\bf Appendix B: Explicit presentation of the new supersymmetric $D$-module reps}}\par
~\par
We present here, for completeness, the explicit constructions of the new $D$-module reps introduced in Section {\bf 7}. They are\par
{\em i}) the
 {\em inhomogeneous} $D$-module reps of the finite $d=1$ superconformal algebras
$osp(1|2)$, $sl(2|1)$, $B(1,1)$, $A(1,1)$ and\par
{\em ii}) the (both {\em homogeneous} and {\em inhomogeneous}) $D$-module reps of the centerless
${\cal N}=1,2,3,4$ superVirasoro algebras.\par
The $D$-module reps with ${\cal N}=1,2,4$ act on the $({\cal N}+1|{\cal N})$ supermultiplets $m$, \\
$m^T =(\varphi_1,\ldots,\varphi_k, g_1,\ldots, g_{{\cal N}-k}, 1| \psi_1,\ldots,\psi_{\cal N})$,
with component fields $\varphi_a, g_i,\psi_\alpha$ and constant entry $1$ in the $({\cal N}+1)$-th position.
The ${\cal N}=3$ $D$-module rep acts on a $(5|4)$ supermultiplet with $1$ in the $5$-th position. The homogeneous $D$-module reps are recovered by deleting the row and the column associated with the constant entry $1$ in the supermultiplet.\par
For ${\cal N}=1$, in matrix form and in the hyperbolic presentation, we can write for the centerless superVirasoro generators
\bea
 Q^0_r &=& e^{rt}\left(\begin{array}{ccc} 0&0&1\\
0&0&0\\-\partial_t-2r\lambda&-2r\rho&0\end{array}\right),\nonumber\\
L_n&=& e^{nt}\left(\begin{array}{ccc} -\partial_t-n\lambda&-n\rho&0\\
0&0&0\\0&0&-\partial_t-\frac{1}{2}n(1+2\lambda)\end{array}\right).\nonumber\\
\eea
The inhomogeneous $D$-module rep of $osp(1|2)$ is recovered for $n=0,\pm1$, $r=\pm\frac{1}{2}$ and by setting $\lambda=0$.\par
To save space, in the remaining cases we limit to present here the odd generators (the even generators are recovered, see (\ref{svir}), from their anticommutators) and write them in terms of the $E_{ij}$ matrices, whose entries are $1$ in the $i$-th row, $j$-th column and vanishing otherwise.
We have,
\par
for ${\cal N}=2$:
\\
the $(2,2,0)$ rep is constructed from
\bea
 Q^0_r &=& e^{rt} [E_{14} + E_{25} - 2(E_{43} + E_{53})r\rho - (E_{41} + E_{52})(\partial_t + 2r\lambda)], \nonumber\\
 Q^1_r &=& e^{rt} [E_{15} - E_{24} + 2(E_{43}-E_{53})r\rho + (E_{42} -E_{51})(\partial_t + 2r\lambda)];
\eea
the $(1,2,1)$ rep is constructed from
\bea
Q^0_r &=& e^{rt} [E_{14} + E_{52} - 2E_{43}r\rho - E_{25}r - (E_{25} + E_{41})(\partial_t + 2r\lambda)], \nonumber\\
Q^1_r &=& e^{rt} [E_{15} - E_{42} -2E_{53}r\rho + E_{24}r + (E_{24} - E_{51})(\partial_t+2r\lambda)];
\eea
the $(0,2,2)$ rep is constructed from
\bea
Q^0_r &=& e^{rt} [E_{41} + E_{52} - (E_{14} +  E_{25})(\partial_t + r +2r\lambda)], \nonumber\\
Q^1_r &=& e^{rt} [E_{51} - E_{42} + (E_{24} - E_{15})(\partial_t + r + 2r\lambda)];
\eea

for ${\cal N}=4$:
\\
the $(4,4,0)$ rep is constructed from
\bea
Q^0_r &=& e^{rt} [E_{16} + E_{27} + E_{38} + E_{49} - 2(E_{65} + E_{75} + E_{85} + E_{95})r\rho\nonumber\\&& - (E_{61} + E_{72} + E_{83} + E_{94})(\partial_t + 2r\lambda)], \nonumber\\
Q^1_r &=& e^{rt} [E_{17} - E_{26} - E_{39} + E_{48} + 2(E_{65} - E_{75} - E_{85} + E_{95})r\rho \nonumber\\&& + (E_{62} - E_{71} - E_{84} + E_{93})(\partial_t + 2r\lambda)], \nonumber\\
Q^2_r &=& e^{rt} [E_{18} + E_{29} - E_{36} - E_{47} + 2(E_{65} + E_{75} - E_{85} - E_{95})r\rho\nonumber\\&& + (E_{63} + E_{74} - E_{81} - E_{92})(\partial_t + 2r\lambda)], \nonumber\\
Q^3_r &=& e^{rt} [E_{19} - E_{28} + E_{37} - E_{46} + 2(E_{65} - E_{75} + E_{85} - E_{95})r\rho \nonumber\\&& + (E_{64} - E_{73} + E_{82} - E_{91})(\partial_t + 2r\lambda)];
\eea
the $(3,4,1)$ rep is constructed from
\bea
Q^0_r &=& e^{rt} [E_{16} + E_{27} + E_{38} + E_{94} - 2(E_{65} + E_{75} + E_{85})r\rho - E_{49}r \nonumber\\&& - (E_{49} + E_{61} + E_{72} + E_{83})(\partial_t + 2r\lambda)], \nonumber\\
Q^1_r &=& e^{rt} [E_{17} - E_{26} - E_{39} + E_{84} + 2(E_{65} - E_{75} + E_{95})r\rho - E_{48}r \nonumber\\&&+ (E_{62} - E_{48} - E_{71} + E_{93})(\partial_t + 2r\lambda)], \nonumber\\
Q^2_r &=& e^{rt} [E_{18} + E_{29} - E_{36} - E_{74} + 2(E_{65} - E_{85} - E_{95})r\rho + E_{47}r \nonumber\\&& + (E_{47} + E_{63} - E_{81} - E_{92})(\partial_t + 2r\lambda)], \nonumber\\
Q^3_r &=& e^{rt} [E_{19} - E_{28} + E_{37} - E_{64} + 2(E_{85} - E_{75} - E_{95})r\rho + E_{46}r\nonumber\\&& + (E_{46} - E_{73} + E_{82} - E_{91})(\partial_t + 2r\lambda)];
\eea
the $(2,4,2)$ rep is constructed from
\bea
Q^0_r &=& e^{rt} [E_{16} + E_{27} + E_{83} + E_{94} - 2(E_{65} + E_{75})r\rho - (E_{38} + E_{49})r \nonumber\\&&- (E_{38} + E_{49} + E_{61} + E_{72})(\partial_t + 2r\lambda)], \nonumber\\
Q^1_r &=& e^{rt} [E_{17} - E_{26} + E_{84} - E_{93} + 2(E_{65} - E_{75})r\rho + (E_{39} - E_{48})r\nonumber\\&& + (E_{39} - E_{48} + E_{62} - E_{71})(\partial_t + 2r\lambda)], \nonumber\\
Q^2_r &=& e^{rt} [E_{18} + E_{29} - E_{63} - E_{74} - 2(E_{85} + E_{95})r\rho + (E_{36} + E_{47})r \nonumber\\&&+ (E_{36} + E_{47} - E_{81} - E_{92})(\partial_t + 2r\lambda)], \nonumber\\
Q^3_r &=& e^{rt} [E_{19} - E_{28} - E_{64} + E_{73} + 2(E_{85} - E_{95})r\rho + (E_{46} - E_{37})r \nonumber\\&&+ (E_{46} - E_{37} + E_{82} - E_{91})(\partial_t + 2r\lambda)];
\eea
the $(1,4,3)$ rep is constructed from
\bea
Q^0_r &=& e^{rt} [E_{16} + E_{72} + E_{83} + E_{94} - 2E_{65}r\rho\nonumber\\&& - (E_{27} + E_{38} + E_{49})r - (E_{27} + E_{38} + E_{49} + E_{61})(\partial_t + 2r\lambda)], \nonumber\\
Q^1_r &=& e^{rt} [E_{17} - E_{62} + E_{84} - E_{93} - 2E_{75}r\rho + (E_{26} + E_{39} - E_{48})r \nonumber\\&&+ (E_{26} + E_{39} - E_{48} - E_{71})(\partial_t + 2r\lambda)], \nonumber\\
Q^2_r &=& e^{rt} [E_{18} - E_{63} - E_{74} + E_{92} - 2E_{85}r\rho + (E_{36} - E_{29} + E_{47})r \nonumber\\&&+ (E_{36} - E_{29} + E_{47} - E_{81})(\partial_t + 2r\lambda)], \nonumber\\
Q^3_r &=& e^{rt} [E_{19} - E_{64} + E_{73} - E_{82} - 2E_{95}r\rho + (E_{28} - E_{37} + E_{46})r\nonumber\\&& + (E_{28} - E_{37} + E_{46} - E_{91})(\partial_t + 2r\lambda)];
\eea
the $(0,4,4)$ rep is constructed from
\bea
Q^0_r &=& e^{rt} [E_{61} + E_{72} + E_{83} + E_{94} - (E_{16} + E_{27} + E_{38} \nonumber\\&&+ E_{49})(\partial_t + r + 2r\lambda)], \nonumber\\
Q^1_r &=& e^{rt} [E_{71} - E_{62} + E_{84} - E_{93}\nonumber\\&& + (E_{26} - E_{17} + E_{39} - E_{48})(\partial_t + r + 2r\lambda)], \nonumber\\
Q^2_r &=& e^{rt} [E_{81} - E_{63} - E_{74} + E_{92} \nonumber\\&&+ (E_{36} - E_{18} - E_{29} + E_{47})(\partial_t + r + 2r\lambda)], \nonumber\\
Q^3_r &=& e^{rt} [E_{73} - E_{64} - E_{82} + E_{91} \nonumber\\&&+ (E_{28} - E_{19} - E_{37} + E_{46})(\partial_t + r + 2r\lambda)].
\eea 
The above operators produce $D$-module reps for the respective superalgebras only at the critical values,
which have been presented in Section {\bf 7}, for $\lambda$ and $\rho$. The inhomogeneous $D$-module reps of the finite $d=1$ superconformal algebras are obtained for $r=\pm\frac{1}{2}$ and $\lambda=0$.\par
The ${\cal N}=3$ $D$-module rep of the centerless superVirasoro algebra (\ref{n3}) exists for arbitrary values
of $\lambda$ and $\rho$. At $\lambda=0$ the restriction to the $B(1,1)$ subalgebra generators produces the $(1,3,3,1)_{0,\rho}$ inhomogeneous $D$-module rep of $B(1,1)$.\par
To reconstruct the full $D$-module rep is sufficient to present the three $Q^i$'s operators.\par
The ${\cal N}=3$ SuperVirasoro $(1,3,3,1)$ rep is obtained from
\bea
Q^1_r &=& e^{rt} [E_{17} - E_{39} - E_{62} + E_{84} - 2E_{75}r\rho\nonumber\\&& + (E_{26} - E_{48} + 2E_{93})r + (E_{26} - E_{48} - E_{71} + E_{93})(\partial_t + 2r\lambda)], \nonumber\\
Q^2_r &=& e^{rt} [E_{18} + E_{29} - E_{63} - E_{74} - 2E_{85}r\rho \nonumber\\&&+ (E_{36} + E_{47} - 2E_{92})r + (E_{36} + E_{47} - E_{81} - E_{92})(\partial_t + 2r\lambda)], \nonumber\\
Q^3_r &=& e^{rt} [E_{16} + E_{49} + E_{72} + E_{83} - 2E_{65}r\rho - (E_{27} + E_{38} + 2E_{94})r \nonumber\\&&- (E_{27} + E_{38} + E_{61} + E_{94})(\partial_t + 2r\lambda)].
\eea

{~}\par
\renewcommand{\theequation}{C.\arabic{equation}}
\setcounter{equation}{0}
 
{\Large{\bf Appendix C: Homogeneous versus Inhomogeneous actions: the ${\cal N}=4$ cases revisited}}\par
~\par
The subtle issues of the relation of the homogeneous versus inhomogeneous actions is already encountered 
in systems with ${\cal N}=0,1,2$ supersymmetries. It is, however, of particular relevance for ${\cal N}=4$ models due to the criticality of the scaling dimension $\lambda$. Indeed, for the $(1,4,3)$ superconformal actions of Section {\bf 9}, $\lambda$ coincides with $\alpha$, the parameter specifying the superconformal symmetry algebra $D(2,1;\alpha)$. \par
One should note that the homogeneous actions (\ref{n4hompar}) and (\ref{n4homhyp}) are not defined at $\alpha =0$ and that even their $\alpha\rightarrow 0$ limit produces a trivial vanishing Lagrangian ($\lim_{\alpha\rightarrow 0}{\cal L}(\alpha)=0$) if the normalization factor $C$ is kept constant. On the other hand, the inhomogeneous actions (\ref{n4inhpar}) and (\ref{n4inhhyp}) are defined at $\lambda=0$ (therefore, at $\alpha=0$).\par
In order to facilitate the comparison of the different actions it is convenient to bring them into a standard form, with the help of fields redefinitions. For this purpose we require the kinetic part of the action being expressed, in terms of the new fields, (denoted as ${\overline\varphi}, {\overline \psi}_I, {\overline g}_i$, with $i=1,2,3$ and $I=0,1,2,3$) as a constant kinetic term. For this reason the new basis of fields will be called
the ``constant kinetic basis".\par
For the two homogeneous (both parabolic and hyperbolic) actions, the fields redefinition is accomplished by the transformations
\bea \label{newhombasis}
\overline{\phi} &=& -2\alpha\phi^{-\frac{1}{2\alpha}}, \nonumber\\
\overline{\psi}_I &=& \phi^{-\frac{1+2\alpha}{2\alpha}}\psi_I, \nonumber\\
\overline{g}_i &=& \phi^{-\frac{1+2\alpha}{2\alpha}} g_i. 
\eea 

For the two inhomogeneous (both parabolic and hyperbolic) actions, the fields redefinition is given by the transformations
\bea \label{newinhbasis}
\overline{\phi} &=& -2\rho e^{-\frac{\phi}{2\rho}}, \nonumber\\
\overline{\psi}_I &=& e^{-\frac{\phi}{2\rho}}\psi_I, \nonumber\\
\overline{g}_i &=&e^{-\frac{\phi}{2\rho}}g_i. 
\eea

The actions (\ref{n4hompar}), (\ref{n4homhyp}), (\ref{n4inhpar}), (\ref{n4inhhyp}), in their respective ``constant kinetic basis", are expressed by

\begin{enumerate}
\item[i)] \emph{Homogeneous parabolic case:}
\bea\label{newn4hompar}
{\cal L}&=& C(\dot{\bar{\phi}}^2 + \bar{\psi}_I\dot{\bar{\psi}}_I+\bar{g_i}^2) + \frac{2(1+2\alpha)C}{\bar{\phi}}\Big(\bar{\psi_0}\bar{\psi_i}\bar{g_i} + \frac{1}{2}\epsilon^{ijk}\bar{\psi}_i\bar{\psi}_j\bar{g}_k\Big) + \nonumber\\
&&\frac{2(1+2\alpha)(1+3\alpha)C}{3\bar{\phi}^2}\epsilon^{ijk}\bar{\psi}_0\bar{\psi}_i\bar{\psi}_j\bar{\psi}_k,
\eea

\item[ii)] \emph{Homogeneous hyperbolic case:}
\bea\label{newn4homhyp}
{\cal L}&=& C(\dot{\bar{\phi}}^2 + \mu\bar{\psi}_I\dot{\bar{\psi}}_I+ \mu^2\bar{g_i}^2) + \frac{2(1+2\alpha)\mu^2C}{\bar{\phi}}\Big(\bar{\psi_0}\bar{\psi_i}\bar{g_i} + \frac{1}{2}\epsilon^{ijk}\bar{\psi}_i\bar{\psi}_j\bar{g}_k\Big)+ \nonumber\\&& \frac{2(1+2\alpha)(1+3\alpha)\mu^2C}{3\bar{\phi}^2}\epsilon^{ijk}\bar{\psi}_0\bar{\psi}_i\bar{\psi}_j\bar{\psi}_k  + \frac{\mu^2C}{4}\bar{\phi}^2, 
\eea

\item[iii)] \emph{Inhomogeneous parabolic case:}
\bea\label{newn4inhpar}
{\cal L}&=& C(\dot{\bar{\phi}}^2 + \bar{\psi}_I\dot{\bar{\psi}}_I+\bar{g_i}^2) + \frac{2C}{\bar{\phi}}\Big(\bar{\psi_0}\bar{\psi_i}\bar{g_i} + \frac{1}{2}\epsilon^{ijk}\bar{\psi}_i\bar{\psi}_j\bar{g}_k\Big) + \nonumber\\
&&\frac{2C}{3\bar{\phi}^2}\epsilon^{ijk}\bar{\psi}_0\bar{\psi}_i\bar{\psi}_j\bar{\psi}_k, 
\eea

\item[iv)] \emph{Inhomogeneous hyperbolic case:}
\bea\label{newn4inhhyp}
{\cal L}&=& C(\dot{\bar{\phi}}^2 + \mu\bar{\psi}_I\dot{\bar{\psi}}_I+ \mu^2\bar{g_i}^2) + \frac{2\mu^2C}{\bar{\phi}}\Big(\bar{\psi_0}\bar{\psi_i}\bar{g_i} + \frac{1}{2}\epsilon^{ijk}\bar{\psi}_i\bar{\psi}_j\bar{g}_k\Big) +\nonumber\\ &&\frac{2\mu^2C}{3\bar{\phi}^2}\epsilon^{ijk}\bar{\psi}_0\bar{\psi}_i\bar{\psi}_j\bar{\psi}_k  + \frac{\mu^2C}{4}\bar{\phi}^2.
\eea
\end{enumerate}

We see that, in this new basis, the $\alpha\rightarrow 0$ limit of the homogeneous actions are well-defined and coincide with the actions obtained from the inhomogeneous transformations. The extra potential term
of the hyperbolic case is an oscillator potential.\par
In the constant kinetic basis the superconformal transformations are realized non-linearly.
The actions of the eight odd generators (the whole superconformal algebra is recovered through their repeated anticommutators) are given by
(here $r=\pm{\frac{1}{2}}$; in the parabolic case $r=-\frac{1}{2}$ corresponds to the four supersymmetry transformations, while $r=\frac{1}{2}$ corresponds
to their four superconformal partners):\\
- in the parabolic case (here $f_r = t^{r-\frac{1}{2}}$):
\bea \label{nlinpar}
Q^i_r \bphi &=& tf_r \bpsi_i, \nonumber\\
Q^i_r \bg_j &=& f_r\epsilon^{ijk}[t\dot{\bpsi}_k + (2r+1)(\alpha+\frac{1}{2})\bpsi_k ] + f_r\delta^{ij}[t\dot{\bpsi} + (2r+1)(\alpha+\frac{1}{2})\bpsi] + \nonumber\\ 
&& (1+2\alpha)\frac{tf_r}{\bphi}[\bpsi_i\bg_j - \epsilon^{ijk}\dot{\bphi}\bpsi_k - \delta^{ij}\dot{\bphi}\bpsi], \nonumber\\
Q^i_r \bpsi &=& -tf_r\bg_i - (1+2\alpha) tf_r\frac{\bpsi\bpsi_i}{\bphi}, \nonumber\\
Q^i_r \bpsi_j &=& tf_r\epsilon^{ijk}\bg_k - f_r\delta^{ij}\big[t\dot{\bphi} + (2r+1)\alpha\bphi\big] + (1+2\alpha)f_r[t\frac{\bpsi_i\bpsi_j}{\bphi} + \delta^{ij}(r+\frac{1}{2})\bphi], \nonumber\\
Q^0_r \bphi &=& tf_r\bpsi, \nonumber\\
Q^0_r \bg_i &=& -f_r[t\dot{\bpsi}_i + (2r+1)(\alpha+\frac{1}{2})\bpsi_i] + (1+2\alpha)\frac{tf_r}{\bphi}[\dot{\bphi}\bpsi_i + \bpsi\bg_i], \nonumber\\
Q^0_r \bpsi &=& -f_r[t\dot{\bphi} + (2r+1)\alpha\bphi] + (1+2\alpha)f_r(r+\frac{1}{2})\bphi, \nonumber\\
Q^0_r \bpsi_i &=& tf_r\bg_i + (1+2\alpha)tf_r\frac{\bpsi\bpsi_i}{\bphi}.
\eea
- in the hyperbolic case (here $f_r = \frac{e^{\mu rt}}{\mu}$):
\bea \label{nlinhyp}
Q^i_r \bphi &=& \mu f_r \bpsi_i, \nonumber\\
Q^i_r \bg_j &=& f_r\epsilon^{ijk}[\dot{\bpsi}_k + 2r\mu(\alpha+\frac{1}{2})\bpsi_k ] + f_r\delta^{ij}[\dot{\bpsi} + 2r\mu(\alpha+\frac{1}{2})\bpsi] + \nonumber\\ 
&& (1+2\alpha)\frac{f_r}{\bphi}[\mu\bpsi_i\bg_j - \epsilon^{ijk}\dot{\bphi}\bpsi_k - \delta^{ij}\dot{\bphi}\bpsi], \nonumber\\
Q^i_r \bpsi &=& -\mu f_r\bg_i - (1+2\alpha) \mu f_r\frac{\bpsi\bpsi_i}{\bphi}, \nonumber\\
Q^i_r \bpsi_j &=& \mu f_r\epsilon^{ijk}\bg_k - f_r\delta^{ij}[\dot{\bphi} + 2r\mu\alpha\bphi] + (1+2\alpha)\mu f_r[\frac{\bpsi_i\bpsi_j}{\bphi} + \delta^{ij}r\bphi], \nonumber\\
Q^0_r \bphi &=& \mu f_r\bpsi, \nonumber\\
Q^0_r \bg_i &=& -f_r[\dot{\bpsi}_i + 2r\mu(\alpha+\frac{1}{2})\bpsi_i] + (1+2\alpha)\frac{f_r}{\bphi}[\dot{\bphi}\bpsi_i + \mu\bpsi\bg_i], \nonumber\\
Q^0_r \bpsi &=& -f_r[\dot{\bphi} + 2r\mu\alpha\bphi] + (1+2\alpha)\mu f_rr\bphi, \nonumber\\
Q^0_r \bpsi_i &=& \mu f_r\bg_i + (1+2\alpha)\mu f_r\frac{\bpsi\bpsi_i}{\bphi}.
\eea
The superconformal transformations corresponding to the inhomogeneous cases are recovered by setting $\alpha=0$.\par
Some comments are in order. Contrary to the ``linear supersymmetry basis of fields" discussed in Section {\bf 9}, the inhomogeneous actions are now recovered as a non-singular $\alpha\rightarrow 0$ limit. 
On the other hand, the linearization of the supersymmetry transformations (given by the inverse of formula (\ref{newinhbasis})), is not recovered from the inverse of (\ref{newhombasis}) in the $\alpha\rightarrow 0$ limit. In other words, the linearization of the $\alpha=0$ supersymmetry (\ref{nlinpar},\ref{nlinhyp}) requires the new notion of inhomogeneous superconformal transformations.\par
The special $\alpha=-\frac{1}{2}$ point implies that the Lagrangian coincides with a free kinetic term (with the addition of the oscillator potential in the hyperbolic case). At this special point the ``constant kinetic basis" coincides with the ``linear supersymmetry basis". \par
This makes clear why we had to discuss the generalization of the Papadopoulos construction for arbitrary values of $\lambda$ in the first place. To obtain the actions (\ref{newn4hompar},
\ref{newn4homhyp},\ref{newn4inhpar},\ref{newn4inhhyp}) based on the non-linear superconformal transformations (\ref{nlinpar},\ref{nlinhyp}) we formulated at first  the linear superconformal problem (for arbitrary $\lambda$'s). In the linear case the situation is under control and, for low values of ${\cal N}$, the linear superconformal transformations are classified.

\par
{~}\par
\renewcommand{\theequation}{D.\arabic{equation}}
\setcounter{equation}{0}
 
{\Large{\bf Appendix D: Note on the hyperbolic/trigonometric superconformally invariant one-dimensional models}}\par
~\par

It is worth pointing out explicitly that the one-dimensional superconformally invariant hyperbolic or trigonometric theories (contrary to their parabolic counterparts) discussed in this paper are not supersymmetric theories (at least if we assume the ordinary
sense of the word ``supersymmetry").\par
The ordinary supersymmetry requires, for a given ${\cal N}$, that a set of ${\cal N}$ fermionic
symmetry generators $Q_i$ closes the supersymmetry algebra $\{Q_i,Q_j\}=2\delta_{ij}H$, $[H,Q_i]=0$ ($i,j=1,\ldots, {\cal N}$), where $H$ is the time-derivative operator (the ``Hamiltonian").\par
In the hyperbolic/trigonometric cases, ${\cal N}$ fermionic symmetry generators can be found. They are the square roots of a symmetry generator (let's call it $Z$), which does not
coincide with the Hamiltonian $H$. As a matter of fact, in the hyperbolic/trigonometric cases, two independent symmetry subalgebras $\{Q_i^\pm,Q_j^\pm\}=2\delta_{ij}Z^\pm$, $[Z^\pm,Q_i^\pm]=0$ (with
$Z^+\neq H$ and $Z^-\neq H$) are encountered. In the parabolic cases two independent symmetry subalgebras are also encountered and one of them can be identified with the
ordinary supersymmetry ($Z^-= H$, $Z^+\neq H$).\par
In the hyperbolic/trigonometric cases the Hamiltonian $H$ continues to be a symmetry operator. It belongs, however, to the $0$-grading sector of the superconformal algebra and {\em is not}
the square of any fermionic symmetry operator (contrary to the operators $Z^\pm$, which belong to the $\pm 1$ grading sectors, respectively).\par
These points can be illustrated with the simplest hyperbolic example, the ${\cal N}=1$ theory based on the $(1,1)$ supermultiplet (a single bosonic field $\varphi$ and a single fermionic field $\psi$) admitting constant kinetic term and $osp(1|2)$ invariance. This hyperbolic action can be written as
\bea\label{hypern1}
{\cal S} &=& \int dt ({\dot\varphi}^2 -\psi{\dot \psi}+\varphi^2).
\eea
The five invariant operators (closing the $osp(1|2)$ algebra) are given by
\bea
Q^{\pm} \varphi = e^{\pm t} \psi, && Q^\pm \psi = e^{\pm t} ({\dot\varphi}\mp \varphi),\nonumber\\
Z^\pm \varphi =e^{\pm 2 t} ({\dot\varphi}\mp \varphi), && Z^\pm \psi = e^{\pm 2t}{\dot \psi},\nonumber\\
H \varphi ={\dot\varphi}, && H\psi = {\dot \psi}.
\eea
One shoulde note that $Z^\pm=({Q^\pm})^2$.\par
No change  of time variable $t\mapsto \tau (t)$ allows to represent either $Z^+$ or $Z^-$ as a time-derivative operator with respect to the new time $\tau$.
\par
The same considerations immediately apply to the other hyperbolic cases with ${\cal N}>1$ and to all the trigonometric cases (possessing a well-defined, bounded below, potential) which, as discussed in Section {\bf 2}, can be recovered from their hyperbolic counterparts.\par
It could be perhaps useful to employ the notion of {\em weak supersymmetry} to describe the nature of systems that, like (\ref{hypern1}), possess a symmetry invariance $Z$ which is the square of fermionic symmetry operators and such that (in contrast with the ordinary, {\em strong} supersymmetry) $Z\neq H$, $H$ being the Hamiltonian.  The notion of ``weak supersymmetry", in a different but related context, has already been introduced in \cite{smi} (see also
\cite{{ivsi},{ivsi2}}).
\\ {~}~
\par {\Large{\bf Acknowledgments}}
{}~\par{}~\par
We are grateful to S. Khodaee and Z. Kuznetsova  for clarifying discussions. We are grateful to the Referee for useful insights and for motivating us to add Appendix {\bf C}.\par The work received support from CNPq.

\end{document}